# Biseparability of 3-qubits density matrices using Hilbert-Schmidt decompositions: Sufficient conditions and explicit expressions

Y. Ben-Aryeh[*] and A. Mann[†]

*Physics Department, Technion-Israel Institute of Technology,*

*Haifa 32000, Israel*

[*] $phr65yb@physics.technion.ac.il$ ;  [†] $ady@physics.technion.ac.il$

**Abstract**

Hilbert-Schmidt (HS) decompositions and Frobenius norms are used to analyze biseparability of 3-qubit systems, with particular emphasis on density matrices with maximally disordered subsystems (MDS) and on the $W$ state mixed with white noise. The biseparable form of a MDS density matrix is obtained by using the Bell states of a 2-qubit subsystem, multiplied by density matrices of the third qubit, which include the relevant HS parameters. Using our methods a sufficient condition and explicit biseparability of the $W$ state mixed with white noise are given. They are compared with the sufficient condition for explicit full separability given in a previous work.

Condensed paper title: Biseparability of 3 qubits.

## 1. Introduction

In previous works [1-4] we found that the Hilbert-Schmidt (HS) decomposition of a density matrix is a very useful method for analyzing various properties of n-qubit systems related to quantum information and computation (see reviews [5-6], books [7-11]). We showed [3,4] that a <u>sufficient</u> condition for full separability is given by the $l_1$ norm [12] of the HS parameters, i.e. sum of the absolute values of those parameters is bounded by 1. The $l_1$ norm is not invariant under unitary transformations, and we found methods to decrease it by local unitary transformations including singular value decompositions (SVD) [13,14] (see Ref. 4, Sec. 2.3).

For 2 qubits the Peres-Horodecki (PH) criterion of partial transpose (PT) [15,16] is necessary and sufficient for entanglement/seperability. For more than 2 qubits the situation is much more complicated. If the PH criterion yields a negative eigenvalue then we know that the given $\rho$ is not fully separable.

However, if the PT yields a valid density matrix no information is gained (see Ref.4, Sec. 2.2, Case A). For 3 qubits one should distinguish between full separability, biseparability and genuine entanglement.[17] Sufficient conditions for full separability of 3-qubits were analyzed by us in previous work.[4] In the present work we analyze sufficient conditions for biseparability of 3 qubits. Fulfillment of our criteria implies that the density matrix is not genuinely entangled.

We put much emphasis in the present work on density matrices with maximally disordered subsystems (MDS),[18] i.e. density matrices for which tracing over any subsystem gives the unit density matrix of the remainder. The reason for this is twofold: First: these 3-qubits density matrices have not been studied extensively in the literature and by using our methods we can analyze various properties of such density matrices. Second: we find also that for more general 3-qubit density matrices their HS decompositions include often MDS terms in addition to other terms which have an explicit fully separable form. Therefore the analysis for MDS density matrices is quite useful for studying biseparability for density matrices for which the HS MDS terms are a part of the full density matrix (such analysis is made for the W state mixed with white noise in Sec. 4).

Necessary conditions for biseparablty were given in Ref. 19, equations (2) and (4). Taking into account the positivity of $\rho$ (for example $|\rho_{1,8}| \leq \sqrt{|\rho_{11}||\rho_{88}|}$) it is easy to verify that these necessary conditions are always satisfied for any 3-qubit MDS $\rho$. Here we deal with sufficient conditions.

. The present paper is arranged as follows:

In Sections (2-3) we find explicit biseparable expressions for MDS density matrices. It has been shown that for odd number of qubits with MDS the PH criterion does not give any information about entanglement (see Ref. 4, Eq. (2.16)). In Sec. 2 we show explicit biseparability of a very simple MDS density matrix with special 3 HS parameters. In this example of one triad of HS parameters, biseparability (say of qubit A with respect to the other two qubits BC) is achieved by representing the BC system by its Bell states.[20] The condition for biseparability is that the Frobenius norm[13] of the HS parameters is bounded by 1. In this special example, this condition is fulfilled since it is the condition that the eigenvalues of the given matrix are non-negative. Similar expressions can be obtained for biseparability of B with respect to AC or biseparability of C with respect to AB.

Based on the special density matrix of Section 2, we show in Sec.3 that the 27 HS parameters may be grouped into 9 triads, related to the triad of the special example by local unitary transformations. Each triad (with the unit matrix) by itself is a biseparable density matrix, since the Frobenius norm of its 3

HS parameters is bounded by 1, which is the necessary condition that it is a density matrix. For more than one triad, we show that a sufficient condition for biseparability is that the sum of their Frobenius norms is bounded by 1. In some special examples of up to 3 triads, this condition was shown to be always fulfilled, as it is a necessary condition that it is a density matrix.

In Sec. 4 we discuss the explicit biseparability of the $W$ state mixed with white noise. Using the methods of the previous Sections we find that when the probability $p$ of the $W$ state is less than 0.1937 it can be written explicitly in a biseparable form. This should be compared with our previous result [4] that for $p < 0.1111$ it may be written explicitly in a fully separable form.

## 2. Explicit biseparability of a simple MDS density matrix

We analyze here the conditions for biseparability of the following 3-qubit very simple MDS density matrix:

$$8\rho_1 = (I)_A \otimes (I)_B \otimes (I)_C + \\ R_{111}(\sigma_x)_A \otimes (\sigma_x)_B \otimes (\sigma_x)_C + R_{222}(\sigma_y)_A \otimes (\sigma_y)_B \otimes (\sigma_y)_C + R_{333}(\sigma_z)_A \otimes (\sigma_z)_B \otimes (\sigma_z)_C \quad (2.1)$$

Here, $I$ represents the unit $2 \times 2$ matrix, the subscripts A,B,C refer to the three qubits, $\sigma_x, \sigma_y, \sigma_z$ are the three Pauli matrices, $\otimes$ denotes outer product, and $R_{111}$, $R_{222}$, and $R_{333}$ are real parameters. The 8 eigenvalues of this density matrix are given by

$$\lambda_1 = \lambda_2 = \lambda_3 = \lambda_4 = (1/8)\left[1 + \sqrt{\sum_{i=1}^{3} R_{iii}^2}\right] \;;$$
$$\lambda_5 = \lambda_6 = \lambda_7 = \lambda_8 = (1/8)\left[1 - \sqrt{\sum_{i=1}^{3} R_{iii}^2}\right] \quad (2.2)$$

Hence it is a density matrix when $R_{iii}$ are within the unit sphere, i.e. when

$$\sum_{i=1}^{3} R_{iii}^2 \leq 1 \quad . \quad (2.3)$$

The sufficient condition for full separability [4] is given by

$$\sum_{i=1}^{3}|R_{iii}|\leq 1 \quad . \tag{2.4}$$

This condition limits the parameters to be within the appropriate cube inscribed in the unit sphere. Since the condition (2.4) is sufficient but not necessary, the separability/nonseparability of $\rho$ in the volume between the cube and the sphere is not decided by it. We now show that $\rho$ of (2.1) may be written in a biseparable form, and therefore cannot be genuinely entangled.

Explicit biseparable expression for the density matrix (2.1) is given by:

$$8\rho_1 = \begin{cases} \left[(I)_A - R_{111}(\sigma_x)_A + R_{222}(\sigma_y)_A + R_{333}(\sigma_z)_A\right] \otimes \left[\left|\Phi^{(-)}\right\rangle_{BC}\left\langle\Phi^{(-)}\right|_{BC}\right] + \\ \left[(I)_A + R_{111}(\sigma_x)_A - R_{222}(\sigma_y)_A + R_{333}(\sigma_z)_A\right] \otimes \left[\left|\Phi^{(+)}\right\rangle_{BC}\left\langle\Phi^{(+)}\right|_{BC}\right] + \\ \left[(I)_A + R_{111}(\sigma_x)_A + R_{222}(\sigma_y)_A - R_{333}(\sigma_z)_A\right] \otimes \left[\left|\Psi^{(+)}\right\rangle_{BC}\left\langle\Psi^{(+)}\right|_{BC}\right] + \\ \left[(I)_A - R_{111}(\sigma_x)_A - R_{222}(\sigma_y)_A - R_{333}(\sigma_z)_A\right] \otimes \left[\left|\Psi^{(-)}\right\rangle_{BC}\left\langle\psi^{(-)}\right|_{BC}\right] \end{cases} . \tag{2.5}$$

Here $\left|\Phi^{(-)}\right\rangle_{BC}$, $\left|\Phi^{(+)}\right\rangle_{BC}$, $\left|\Psi^{(+)}\right\rangle_{BC}$ and $\left|\Psi^{(-)}\right\rangle_{BC}$ are the Bell states[20] of the qubits pair $B$ and $C$. We used Bell states density matrices which are expanded in terms of Pauli matrices as:

$$4\left|\Phi^{(-)}\right\rangle_{BC}\left\langle\Phi^{(-)}\right|_{BC} = \left[(I)_B \otimes (I)_C - (\sigma_x)_B \otimes (\sigma_x)_C + (\sigma_y)_B \otimes (\sigma_y)_C + (\sigma_z)_B \otimes (\sigma_z)_C\right] \; ;$$

$$4\left|\Phi^{(+)}\right\rangle_{BC}\left\langle\Phi^{(+)}\right|_{BC} = \left[(I)_B \otimes (I)_C + (\sigma_x)_B \otimes (\sigma_x)_C - (\sigma_y)_B \otimes (\sigma_y)_C + (\sigma_z)_B \otimes (\sigma_z)_C\right] \; ;$$

$$4\left|\Psi^{(+)}\right\rangle_{BC}\left\langle\Psi^{(+)}\right|_{BC} = \left[(I)_B \otimes (I)_C + (\sigma_x)_B \otimes (\sigma_x)_C + (\sigma_y)_B \otimes (\sigma_y)_C - (\sigma_z)_B \otimes (\sigma_z)_C\right] \; ; \quad (2.6)$$

$$4\left|\Psi^{(-)}\right\rangle_{BC}\left\langle\Psi^{(-)}\right|_{BC} = \left[(I)_B \otimes (I)_C - (\sigma_x)_B \otimes (\sigma_x)_C - (\sigma_y)_B \otimes (\sigma_y)_C - (\sigma_z)_B \otimes (\sigma_z)_C\right] \; ;$$

Equations (2.1) and (2.5) represent a biseparable density matrix, under the condition

$$\sqrt{R_{111}^2 + R_{222}^2 + R_{333}^2} \leq 1 \quad , \tag{2.7}$$

which is equivalent to the condition (2.3) for Eq. (2.1) to be a density matrix.

Although the above biseparable form is given for the very special density matrix (2.1) we will show in the next Section that by using local unitary transformations a sufficient criterion for a biseparable form may be obtained for more general MDS density matrices.. In this context one should take into account that local unitary transformations of Bell states of qubits $B$ and $C$ will preserve the Bell states properties in the transformed frames of reference. We note that similar expressions of the biseparability (2.5) may be written in terms of the Bell states of $AB$ or $AC$.

## 3. Sufficient condition for biseparability for general 3-qubit MDS density matrix using Bell states

A general 3-qubits MDS density matrix, in the HS decomposition, has the form

$$8\rho = (I)_A \otimes (I)_B \otimes (I)_C + \sum_{l,m,n=1}^{3} R_{l,m,n} (\sigma_l)_A \otimes (\sigma_m)_B \otimes (\sigma_n)_C \equiv (I)_A \otimes (I)_B \otimes (I)_C + R \ . \quad (3.1)$$

We would like to show that

$$\sum_{l,m,n=1}^{3} R_{l,m,n}^2 \leq 1 \quad . \quad (3.2)$$

We note first that

$$Tr(8\rho)^2 = 8 + 8 \sum_{l,m,n=1}^{3} R_{l,m,n}^2 \quad . \quad (3.3)$$

On the other hand

$$Tr(8\rho)^2 = 64 \sum_{i=1}^{8} \lambda_i^2 \quad . \quad (3.4)$$

Here $\lambda_i$ are the 8 eigenvalues of $\rho$. Since $0 \leq \lambda_i \leq 1/4$ (see Ref. 4, comment after Eq. (2.4)) we write (recalling that the 8 $r_i$ come in 4 pairs $\pm |r_i|$ [4]):

$$\lambda_i = \frac{1}{8} + r_i \quad ; \quad |r_i| \leq \frac{1}{8} \quad ; \quad \sum_{i=1}^{8} r_i = 0 \ . \quad (3.5)$$

Hence

$$\sum_{i=1}^{8} \lambda_i^2 = \sum_{i=1}^{8} \left(\frac{1}{8} + r_i\right)^2 = \sum_{i=1}^{8} \left(\frac{1}{64}\right) + \sum_{i=1}^{8} (r_i^2) \leq \frac{1}{8} + \frac{1}{8} = \frac{1}{4} \quad . \quad (3.6)$$

By using Eqs. (3.3-3.6) we get Eq. (3.2). Eq. (3.2) may be generalized to any odd-n MDS density matrix.

Note that the equality in Eq. (3.6) holds only if

$$\lambda_i = \frac{1}{4} \ (i=1,2,3,4) \ ; \ \lambda_j = 0 \ (j=5,6,7,8) \ . \tag{3.7}$$

We note that local unitary transformations, on the qubits $A, B, C$ in Equations. (2.1) and (2.5) will obviously produce other simple biseparable MDS density matrices. For example the density matrix

$$8\rho_2 = (I)_A \otimes (I)_B \otimes (I)_C + \\ R_{132}(\sigma_x)_A \otimes (\sigma_z)_B \otimes (\sigma_y)_C + R_{321}(\sigma_z)_A \otimes (\sigma_y)_B \otimes (\sigma_x)_C + R_{213}(\sigma_y)_A \otimes (\sigma_x)_B \otimes (\sigma_z)_C, \tag{3.8}$$

is obtained from (2.1) by a $90^0$ rotation, of $A$ around $x$, $B$ around $y$, and $C$ around $z$ and inverting the signs of the HS parameters. Therefore in the biseparable form, Eq. (2.5) (and Eq. (2.6)) we have simply to make the following exchanges

$$(\sigma_x)_A \to (\sigma_x)_A \ ; \ (\sigma_y)_A \to (\sigma_z)_A \ ; \ (\sigma_z)_A \to -(\sigma_y)_A \ ; \\ (\sigma_x)_B \to -(\sigma_z)_B \ ; \ (\sigma_y)_B \to (\sigma_y)_B \ ; \ (\sigma_z)_B \to (\sigma_x)_B \ ; \ . \tag{3.9} \\ (\sigma_x)_C \to (\sigma_y)_C \ ; \ (\sigma_y)_C \to -(\sigma_x)_C \ ; \ (\sigma_z)_C \to (\sigma_z)_C$$

and invert the signs of the relevant HS parameters.

Then, by using the corresponding parameters in Eq. (3.1) we obtain $\rho_2$ in the biseparable form:

$$8\rho_2 = \begin{cases} \left[(I)_A - R_{132}(\sigma_x)_A + R_{321}(\sigma_y)_A + R_{213}(\sigma_z)_A\right] \otimes \left[|\tilde{\Phi}^{(-)}\rangle_{BC}\langle\tilde{\Phi}^{(-)}|_{BC}\right] + \\ \left[(I)_A + R_{132}(\sigma_x)_A - R_{321}(\sigma_y)_A + R_{213}(\sigma_z)_A\right] \otimes \left[|\tilde{\Phi}^{(+)}\rangle_{BC}\langle\tilde{\Phi}^{(+)}|_{BC}\right] + \\ \left[(I)_A + R_{132}(\sigma_x)_A + R_{321}(\sigma_y)_A - R_{213}(\sigma_z)_A\right] \otimes \left[|\tilde{\Psi}^{(+)}\rangle_{BC}\langle\tilde{\Psi}^{(+)}|_{BC}\right] + \\ \left[(I)_A - R_{132}(\sigma_x)_A - R_{321}(\sigma_y)_A - R_{213}(\sigma_z)_A\right] \otimes \left[|\tilde{\Psi}^{(-)}\rangle_{BC}\langle\tilde{\psi}^{(-)}|_{BC}\right] \end{cases}. \tag{3.10}$$

The density matrices $|\tilde{\Phi}^{(-)}\rangle_{BC}\langle\tilde{\Phi}^{(-)}|_{BC} ; |\tilde{\Phi}^{(+)}\rangle_{BC}\langle\tilde{\Phi}^{(+)}|_{BC} ; |\tilde{\Psi}^{(+)}\rangle_{BC}\langle\tilde{\Psi}^{(+)}|_{BC} ; |\tilde{\Psi}^{(-)}\rangle_{BC}\langle\tilde{\Psi}^{(-)}|_{BC}$ are obtained from Eq. (2.6) by using the transformations of Eq. (3.9).

Now, suppose we are given a more complicated density matrix, in the form

$$8\rho = (I)_A \otimes (I)_B \otimes (I)_C +$$

$$R_{111}(\sigma_x)_A \otimes (\sigma_x)_B \otimes (\sigma_x)_C + R_{222}(\sigma_y)_A \otimes (\sigma_y)_B \otimes (\sigma_y)_C + R_{333}(\sigma_z)_A \otimes (\sigma_z)_B \otimes (\sigma_z)_C \quad (3.10)$$

$$+ R_{132}(\sigma_x)_A \otimes (\sigma_z)_B \otimes (\sigma_y)_C + R_{321}(\sigma_z)_A \otimes (\sigma_y)_B \otimes (\sigma_x)_C + R_{213}(\sigma_y)_A \otimes (\sigma_x)_B \otimes (\sigma_z)_C$$

We would like to see if it can be written in a biseparable form. For this purpose we note that we can write Eqs. (2.1) in a different form:

$$8\rho_1 \equiv \left(1 - \sqrt{R_{111}^2 + R_{222}^2 + R_{333}^2}\right)\{(I)_A \otimes (I)_B \otimes (I)\}_C + \left(\sqrt{R_{111}^2 + R_{222}^2 + R_{333}^2}\right) 8\tilde{\rho}_1 \quad . \quad (3.11)$$

Here we defined a new density matrix $\tilde{\rho}_1$ given by

$$8\tilde{\rho}_1 = (I)_A \otimes (I)_B \otimes (I)_C + \frac{1}{\sqrt{\sum_{i=1}^{3} R_{iii}^2}} \sum_{i=1}^{3} R_{iii}(\sigma_i)_A \otimes (\sigma_i)_B \otimes (\sigma_i)_C \; ; \sigma_1 \equiv \sigma_x, \sigma_2 \equiv \sigma_y, \sigma_3 \equiv \sigma_z$$

$$(3.12)$$

This density matrix can be written in a biseparable form:

$$8\tilde{\rho}_1 = \left\{\begin{array}{l} \left[(I)_A + \dfrac{-R_{111}(\sigma_x)_A + R_{222}(\sigma_y)_A + R_{333}(\sigma_z)_A}{\left(\sqrt{R_{111}^2 + R_{222}^2 + R_{333}^2}\right)}\right] \otimes \left[\left|\Phi^{(-)}\right\rangle_{BC}\left\langle\Phi^{(-)}\right|_{BC}\right] + \\ \left[(I)_A + \dfrac{R_{111}(\sigma_x)_A - R_{222}(\sigma_y)_A + R_{333}(\sigma_z)_A}{\left(\sqrt{R_{111}^2 + R_{222}^2 + R_{333}^2}\right)}\right] \otimes \left[\left|\Phi^{(+)}\right\rangle_{BC}\left\langle\Phi^{(+)}\right|_{BC}\right] + \\ \left[(I)_A + \dfrac{R_{111}(\sigma_x)_A + R_{222}(\sigma_y)_A - R_{333}(\sigma_z)}{\left(\sqrt{R_{111}^2 + R_{222}^2 + R_{333}^2}\right)}\right] \otimes \left[\left|\Psi^{(+)}\right\rangle_{BC}\left\langle\Psi^{(+)}\right|_{BC}\right] + \\ \left[(I)_A + \dfrac{-R_{111}(\sigma_x)_A - R_{222}(\sigma_y)_A - R_{333}(\sigma_z)_A}{\left(\sqrt{R_{111}^2 + R_{222}^2 + R_{333}^2}\right)}\right] \otimes \left[\left|\Psi^{(-)}\right\rangle_{BC}\left\langle\psi^{(-)}\right|_{BC}\right] \end{array}\right\} . \quad (3.13)$$

Then, using Eq. (3.8) and the transformations (3.9) we get

$$8\rho_2 = \left(1 - \sqrt{R_{132}^2 + R_{321}^2 + R_{213}^2}\right)\{(I)_A \otimes (I)_B \otimes (I)\}_C + \left(\sqrt{R_{132}^2 + R_{321}^2 + R_{333}^2}\right) 8\tilde{\rho}_2 \quad . \quad (3.14)$$

$8\tilde{\rho}_2$ is obtained from Eq. (3.13) by using the transformations of Eq. (3.9) on qubit $A$ and on the Bell states and transforming the HS parameters as:

$$R_{111} \rightarrow R_{132} \quad ; \quad R_{222} \rightarrow R_{321} \quad ; \quad R_{333} \rightarrow R_{213} \quad , \tag{3.15}$$

With these definitions of $\tilde{\rho}_1, \tilde{\rho}_2$ we have for Eq. (3.10) the <u>identity</u>

$$8\rho = (I)_A \otimes (I)_B \otimes (I)_C \left(1 - \sqrt{R_{111}^2 + R_{222}^2 + R_{333}^2} - \sqrt{R_{132}^2 + R_{321}^2 + R_{213}^2}\right) \\ + \left(\sqrt{R_{111}^2 + R_{222}^2 + R_{333}^2}\right) 8\tilde{\rho}_1 + \left(\sqrt{R_{132}^2 + R_{321}^2 + R_{333}^2}\right) 8\tilde{\rho}_2 \tag{3.16}$$

We note that since according to Eq. (3.2) we get $R_{111}^2 + R_{222}^2 + R_{333}^2 + R_{132}^2 + R_{321}^2 + R_{333}^2 \leq 1$, the coefficients of $8\tilde{\rho}_1$ and $8\tilde{\rho}_2$ are smaller than 1. But, for the right hand side to represent a biseparable density matrix we have to require

$$\sqrt{R_{111}^2 + R_{222}^2 + R_{333}^2} + \sqrt{R_{132}^2 + R_{321}^2 + R_{213}^2} \leq 1 \quad . \tag{3.17}$$

Eq. (3.17) represents a sufficient condition for biseparability of the density matrix given by Eqs. (3.10), and (3.16). We note that the condition (3.16) requires that the sum of the Frobenius norms of the two triads of MDS–parameters is not larger than 1.

It is worth noting that in the example for $\rho$ of Eq. (3.10) the sufficient condition for biseparability (3.17) is the necessary condition that $\rho$ is a density matrix. This is so, because the operators in one triad commute with those in the other; therefore the smallest eigenvalue of $8\rho$ is given by $1 - \sqrt{R_{111}^2 + R_{222}^2 + R_{333}^2} - \sqrt{R_{132}^2 + R_{321}^2 + R_{213}^2} \geq 0$.

To treat the general case of MDS density matrix given by (3.1) (up to 27 MDS terms), we can divide $\sum_{l,m,n=1}^{3} R_{l,m,n} (\sigma_x)_A \otimes (\sigma_y)_B \otimes (\sigma_z)_C$ into 9 groups of three triads. Starting with the triad

$$R_{111}(\sigma_x)_A \otimes (\sigma_x)_B \otimes (\sigma_x)_C + R_{222}(\sigma_y)_A \otimes (\sigma_y)_B \otimes (\sigma_y)_C + R_{333}(\sigma_z)_A \otimes (\sigma_z)_B \otimes (\sigma_z)_C$$

of Eq.(2.1), we can apply 8 transformations (similar to those of Eq. (3.9)) to the Pauli matrices of each qubit, obtaining 8 triads with corresponding HS parameters. Each triad may be treated as in Eqs. (3.13, 3.14) to include together with Eq. (3.11) the 27 $R_{l,m,n}$ parameters. Therefore the sufficient condition for biseparability of Eq. (3.1) becomes that the sum of the Frobenius norms of the 9 (at most) triads of MDS-parameters is not larger than 1. Such condition is sufficient for biseparability but the sufficient condition for biseparability may be improved by other methods.

As another simple example we add another triad to $\rho$ of (3.10) to obtain

$$
\begin{aligned}
8\rho = &(I)_A \otimes (I)_B \otimes (I)_C + \\
&R_{111}(\sigma_x)_A \otimes (\sigma_x)_B \otimes (\sigma_x)_C + R_{222}(\sigma_y)_A \otimes (\sigma_y)_B \otimes (\sigma_y)_C + R_{333}(\sigma_z)_A \otimes (\sigma_z)_B \otimes (\sigma_z)_C \\
&+ R_{132}(\sigma_x)_A \otimes (\sigma_z)_B \otimes (\sigma_y)_C + R_{321}(\sigma_z)_A \otimes (\sigma_y)_B \otimes (\sigma_x)_C + R_{213}(\sigma_y)_A \otimes (\sigma_x)_B \otimes (\sigma_z)_C \\
&+ R_{123}(\sigma_x)_A \otimes (\sigma_y)_B \otimes (\sigma_z)_C + R_{312}(\sigma_z)_A \otimes (\sigma_x)_B \otimes (\sigma_y)_C + R_{231}(\sigma_y)_A \otimes (\sigma_z)_B \otimes (\sigma_x)_C
\end{aligned} \quad (3.18)
$$

The sufficient condition for biseparability is

$$\sqrt{R_{111}^2 + R_{222}^2 + R_{333}^2} + \sqrt{R_{132}^2 + R_{321}^2 + R_{213}^2} + \sqrt{R_{123}^2 + R_{312}^2 + R_{231}^2} \leq 1 \quad , \qquad (3.19)$$

Also in this example, the operators in any triad commute with those of the other triads, so that Eq. (3.19) is the necessary condition that the smallest eigenvalue of $\rho$ (Eq. (3.18)) is nonnegative. This does not hold in the general case when the operators in one triad do not commute with those of other triads. In such cases the condition for density matrix does not seem to necessarily imply biseparability.

## 4. Explicit biseparability of the W state mixed with white noise

In the present Section we apply our methods to calculate sufficient conditions for biseparability of $W$ state mixed with white noise. Assuming that the density matrix $\rho(W)$ for the $W$ state with a probability $p$ is mixed with white noise with probability (1-p) we get

$$8\rho(W; mixed) = (1-p)(I)_A \otimes (I)_B \otimes (I)_C + p 8\rho(W) \quad . \qquad (4.1)$$

Here $8\rho(W)$ is given by [4]

$$3 \cdot 8\rho(W) =$$

$$\begin{pmatrix}
0 & 0 & 0 & 0 & 0 & 0 & 0 & 0 \\
0 & 8 & 8 & 0 & 8 & 0 & 0 & 0 \\
0 & 8 & 8 & 0 & 8 & 0 & 0 & 0 \\
0 & 0 & 0 & 0 & 0 & 0 & 0 & 0 \\
0 & 8 & 8 & 0 & 8 & 0 & 0 & 0 \\
0 & 0 & 0 & 0 & 0 & 0 & 0 & 0 \\
0 & 0 & 0 & 0 & 0 & 0 & 0 & 0 \\
0 & 0 & 0 & 0 & 0 & 0 & 0 & 0
\end{pmatrix} \quad (4.2)$$

The HS decomposition of (4.2) is quite complicated and given by

$$\begin{aligned}
3 \cdot 8\rho(W) =\ & 2(\sigma_y)_A \otimes (I)_B \otimes (\sigma_y)_C + 2(I)_A \otimes (\sigma_y)_B \otimes (\sigma_y)_C + 2(\sigma_y)_A \otimes (\sigma_y)_B \otimes (I)_C \\
& -(\sigma_z)_A \otimes (I)_B \otimes (\sigma_z)_C - (I)_A \otimes (\sigma_z)_B \otimes (\sigma_z)_C - (\sigma_z)_A \otimes (\sigma_z)_B \otimes (I)_C \\
& +(\sigma_z)_A \otimes (I)_B \otimes (I)_C + (I)_A \otimes (\sigma_z)_B \otimes (I)_C + (I)_A \otimes (I)_B \otimes (\sigma_z)_C \\
& +2(\sigma_x)_A \otimes (\sigma_z)_B \otimes (\sigma_x)_C + 2(\sigma_z)_A \otimes (\sigma_x)_B \otimes (\sigma_x)_C + 2(\sigma_x)_A \otimes (\sigma_x)_B \otimes (\sigma_z)_C \\
& +2(\sigma_x)_A \otimes (\sigma_x)_B \otimes (I)_C + 2(\sigma_x)_A \otimes (I)_B \otimes (\sigma_x)_C + 2(I)_A \otimes (\sigma_x)_B \otimes (\sigma_x)_C \\
& +2(\sigma_y)_A \otimes (\sigma_y)_B \otimes (\sigma_z)_C + 2(\sigma_y)_A \otimes (\sigma_z)_B \otimes (\sigma_y)_C + 2(\sigma_z)_A \otimes (\sigma_y)_B \otimes (\sigma_y)_C \\
& -3(\sigma_z)_A \otimes (\sigma_z)_B \otimes (\sigma_z)_C + 3(I)_A \otimes (I)_B \otimes (I)_C
\end{aligned} \quad (4.3)$$

In our previous work [4] we have used equations (4.1) and (4.3) and derived the sufficient condition $p \leq 1/9$ for explicit full separability of the density matrix of Eq. (4.1).

We analyze in the present Section a sufficient condition for explicit biseparability for the density matrix of Eq. (4.1) using the HS decomposition of $8\rho(W)$ given by Eq. (4.3). Although in Eq. (4.3) only a part of the density matrix is related to MDS, the use of the methods presented in previous sections can improve the condition for biseparably relative to that of full separability.

The following six MDS terms

$$(\sigma_x)_A \otimes (\sigma_z)_B \otimes (\sigma_x)_C\ ,\ (\sigma_z)_A \otimes (\sigma_x)_B \otimes (\sigma_x)_C\ ,\ (\sigma_x)_A \otimes (\sigma_x)_B \otimes (\sigma_z)_C\ ,$$
$$(\sigma_y)_A \otimes (\sigma_y)_B \otimes (\sigma_z)_C\ ,\ (\sigma_y)_A \otimes (\sigma_z)_B \otimes (\sigma_y)_C\ ,\ (\sigma_z)_A \otimes (\sigma_y)_B \otimes (\sigma_y)_C\ ,$$

can be grouped into 3 pairs:

(a) $\left[(\sigma_x)_A \otimes (\sigma_z)_B \otimes (\sigma_x)_C , (\sigma_y)_A \otimes (\sigma_y)_B \otimes (\sigma_z)_C\right]$

(b) $\left[(\sigma_x)_A \otimes (\sigma_x)_B \otimes (\sigma_z)_C , (\sigma_z)_A \otimes (\sigma_y)_B \otimes (\sigma_y)_C\right]$ .

(c) $\left[(\sigma_z)_A \otimes (\sigma_y)_B \otimes (\sigma_y)_C , (\sigma_y)_A \otimes (\sigma_z)_B \otimes (\sigma_y)_C\right]$

Pair (a) is obtained from the pair $\left[(\sigma_x)_A \otimes (\sigma_x)_B \otimes (\sigma_x)_C , (\sigma_y)_A \otimes (\sigma_y)_B \otimes (\sigma_y)_C\right]$ by the local unitary transformation $U_1$ which rotates $B$ around the $y$ axis, and $C$ around $x$ by $\pi/2$ . Similarly (b) is obtained from the same pair by $U_2$ which rotates $A$ around $x$ and $C$ around y by $\pi/2$, and (c) is obtained by $U_3$ rotating $A$ around $y$ and $B$ around $x$ by $\pi/2$.

Recalling the decomposition of $\rho_1$ (Eq. (2.1) in a biseparable form (Eq. (2.5) ), we get after some tedious calculations the following explicit expression for the biseparability of the density matrix of Eq. (4.1):

$$8\rho(W;mixed) =$$
$$\frac{p}{3}\left\{\begin{array}{l}(I+\sigma_z)_A \otimes (I+\sigma_z)_B \otimes (I-\sigma_z)_C + (I-\sigma_z)_A \otimes (I+\sigma_z)_B \otimes (I+\sigma_z)_C \\ +(I+\sigma_z)_A \otimes (I-\sigma_z)_B \otimes (I+\sigma_z)_C\end{array}\right\}$$
$$+\frac{p}{3}\left\{(I+\sigma_Y)_A \otimes (I+\sigma_Y)_B \otimes (I+\sigma_Y)_C + (I-\sigma_Y)_A \otimes (I-\sigma_Y)_B \otimes (I-\sigma_z)_C\right\}$$
$$+\frac{p}{3}\left\{(I+\sigma_x)_A \otimes (I+\sigma_x)_B \otimes (I+\sigma_x)_C + (I-\sigma_x)_A \otimes (I-\sigma_x)_B \otimes (I-\sigma_x)_C\right\}$$
$$+\frac{2p}{3}\sqrt{2}\sum_{i=1}^{3} U_i \left\{\begin{array}{l}\left[(I)_A + \frac{-(\sigma_x)_A + (\sigma_y)_A}{\sqrt{2}}\right] \otimes \left[\left|\Phi^{(-)}\right\rangle_{BC}\left\langle\Phi^{(-)}\right|_{BC}\right] + \\ \left[(I)_A + \frac{(\sigma_x)_A - (\sigma_y)_A}{\sqrt{2}}\right] \otimes \left[\left|\Phi^{(+)}\right\rangle_{BC}\left\langle\Phi^{(+)}\right|_{BC}\right] + \\ \left[(I)_A + \frac{(\sigma_x)_A + (\sigma_y)_A}{\sqrt{2}}\right] \otimes \left[\left|\Psi^{(+)}\right\rangle_{BC}\left\langle\Psi^{(+)}\right|_{BC}\right] + \\ \left[(I)_A + \frac{-(\sigma_x)_A - (\sigma_y)_A}{\sqrt{2}}\right] \otimes \left[\left|\Psi^{(-)}\right\rangle_{BC}\left\langle\psi^{(-)}\right|_{BC}\right]\end{array}\right\} U_i^{\dagger}$$
$$+\left(1 - \frac{7}{3}p - 2\sqrt{2}\,p\right)(I)_A \otimes (I)_B \otimes (I)_C \qquad (4.4)$$

We get from Eq. (4.4) that for $p \leq \dfrac{1}{7/3 + 2\sqrt{2}} = 0.1937$, $\rho(W;mixed)$ of Eq. (4.1) is not genuinely entangled. This condition can be compared with the condition $p \leq \dfrac{1}{9} \approx 0.1111$ which was obtained by us as a sufficient condition for full separability.[4]

## 6. Summary and discussion

In the present work we treated in Sections 2 and 3 a sufficient condition for biseparability of one qubit (say A) with respect to the other two qubits (BC). This is obtained by using a representation, where the first qubit includes the 3 HS parameters, and it is multiplied by the Bell states of the other two qubits. In Sec. 2 we analyzed the sufficient condition for biseparability of the density matrix (2.1) which includes three HS parameters multiplied by the three diagonal products of Pauli matrices. Explicitly biseparable expression for the density matrix was given in Eqs. (2.5-2.6). A sufficient condition for biseparability for this special case (given by Eq. (2.7)) is that the sum of HS parameters squared will not be larger than 1, which is equivalent to the condition of Eq. (2.1) to be a density matrix. In another form the condition in Eq. (2.7) requires the Frobenius norm[13] of the 3 HS parameters to be not larger than 1.

In Sec. (3) we proved the relation (3.2) showing that for odd-n MDS density matrices the sum of HS parameters squared cannot be larger than 1. The equality with 1 will be obtained only for the special case described by Eq. (3.7). For treating the general case of MDS density matrix given by Eq. (3.1) (up to 27 MDS terms) we can start with the triad of Eq. (2.1) and apply 8 local transformations (similar to those of Eq. (3.9)) to the Pauli matrices of each qubit, obtaining 8 triads with corresponding HS parameters. The sufficient condition for biseparability of Eq. (3.1) becomes then that the sum of the Frobenius norms of the 9 (or less) triads of MDS parameters is not larger than 1. We demonstrated this method for the case in which the 3-qubits MDS density matrix includes two triads of three qubits products given by Eq. (3.10). This density matrix can be written by the explicitly biseparable form given by Eq. (3.16) and the sufficient condition for biseparability in this case is given by Eq. (3.17) requiring that the sum of Frobenius norms of the two triads will not be larger than 1. We demonstrated in Eq. (3.18) a density matrix which includes three triads of Pauli matrices products. The sufficient condition for bisepability is then given by Eq. (3.19) requiring here again that the sum of the three Frobenius norms will not be larger than 1.

We should mention that the present sufficient condition for biseparability does not preclude the possibility that the density matrix is fully separable. In some cases (especially for a large number of HS parameters) it may happen that the sufficient condition for full separability obtained by using SVD [4] is satisfied and then the present condition is not needed.

In Sec. 4, we analyzed explicit biseparability of the density matrix (4.1) for the W state mixed with white noise. In this analysis we used the HS decomposition of the W density matrix where six MDS terms have been grouped into 3 pairs and their biseparable form has been given by the methods analyzed in Sections 2, 3. All other terms in the HS decomposition were given in a fully separable form. We obtained the sufficient condition for biseparability given by $p \leq 0.1937$ in comparison with the sufficient condition for full separability given in our previous work [4] as $p \leq 0.1111$. This result can also be compared with the result obtained by the PH criterion by which for $p > 0.209589$ this density matrix cannot be fully separable. [4] It is interesting to note that for $p > 0.529$ it has been shown [19] that the mixed $W$ state is genuinely entangled.